\documentclass[a4paper,12pt]{article}
\usepackage[dvips]{graphicx}
\usepackage{a4wide}
\usepackage{bbm}% bold math
\usepackage{epsfig} 
\usepackage{amsmath}
\usepackage{color}

\newcommand{\tr}{\,\text{tr}}

\newcommand{\I}{\text{i}}

\newcommand{\fsl}[1]{#1\!\!\!\!/}
\newcommand{\Nf}{N_{\text{f}}}
\newcommand{\Nc}{N_{\text{c}}}

\newcommand{\GQA}{\Gamma_{\text{QA}}}

\definecolor{blue}{rgb}{0,0,1}

\title{\bf Geometry of spin-field coupling \\ on the worldline}
\author{Holger Gies and Jens H\"ammerling \\ \\
     \small \it Institut f\"ur theoretische Physik, Universit\"at Heidelberg
     \\ \small \it Philosophenweg 16, D-69120 Heidelberg, Germany }
 
\begin{document}

\maketitle

$\text{}$

\vspace{-10.3cm} 

{\hfill \small\sf HD-THEP-05-08, 
{  }http://arXiv.org/abs/hep-ph/0505072} 

\vspace{8.5cm}

\begin{abstract}
  We derive a geometric representation of couplings between spin
  degrees of freedom and gauge fields within the worldline approach to
  quantum field theory. We combine the string-inspired methods of the
  worldline formalism with elements of the loop-space approach to gauge
  theory.  In particular, we employ the loop (or area) derivative
  operator on the space of all holonomies which can immediately be
  applied to the worldline representation of the effective action.
  This results in a spin factor that associates the information about
  spin with ``zigzag" motion of the fluctuating field.  Concentrating
  on the case of quantum electrodynamics in external fields, we obtain
  a purely geometric representation of the Pauli term. To one-loop
  order, we confirm our formalism by rederiving the Heisenberg-Euler
  effective action. Furthermore, we give closed-form worldline
  representations for the all-loop order effective action to lowest
  nontrivial order in a small-$\Nf$ expansion.
\end{abstract}

\section{Introduction}

The mapping of quantum field theoretic problems onto the language of
quantum mechanics of point particles in the form of the worldline
formalism \cite{Feynman:1950ir} has become a powerful computational tool
in recent years.  The worldline approach, which can also be viewed as
the field theoretic limit of string theory
\cite{Halpern:1977ru,Fradkin:1985qd,Metsaev:1987ju,Polyakov:ez,%
Bern:1991aq,Strassler:1992zr},
establishes a direct connection between a ``second-quantized'' and a
``first-quantized'' formalism. Particularly for correlators in
background fields, computations simplify drastically with
worldline techniques \cite{Schmidt:1993rk,Schubert:2001he}.

The relation between field theory and quantum particle mechanics can
best be illustrated by the worldline representation of a scalar
field's propagator in Euclidean spacetime,
\begin{equation}
G(x_2,x_1)=\int_0^\infty dT\, e^{-m^2T}\, \mathcal N
\int_{x(0)=x_1}^{x(T)=x_2} \mathcal D x\, e^{-\frac{1}{4} \int_0^T
  d\tau \, \dot x ^2 (\tau)}, \label{eScalProp}
\end{equation}
where the integration parameter $T$ is called propertime, and the path
integral runs over all paths with fixed end points distributed by a
Gau\ss ian velocity weight. The resulting ensemble of paths can be
viewed as the set of possible trajectories of the quantum field. This
associates virtual fluctuations of a field with particle worldlines in
coordinate space, which constitutes a highly intuitive picture for the
nature of quantum fluctuations. Incidentally, the path integral with
Gau\ss ian velocity weight can also be represented by a sum over
trajectories of a random walker \cite{Itzykson:1989sx}.

The standard route to worldline representations of propagators for
higher-spin fields proceeds with the aid of Grassmann-valued path
integrals that encode the spin degrees of freedom as well as the
corresponding algebra \cite{Berezin:1976eg}. Though technically
elegant and computationally powerful, this approach goes along with a
loss of intuition: trajectories in Grassmann-space are difficult to be
visualized. 

An alternative approach has been suggested in
\cite{Strominger:1980xa,Polyakov:1988} for $D=2,3$ dimensions, where
the information about fermionic spin can be encoded in terms of the
``Polyakov spin factor''. This spin factor acts as an
insertion in the path integrand and depends solely on the worldline
itself; for instance, in $D=2$, it can be represented by the trace of
a path-ordered exponential, 
\begin{equation}
\Phi_{\text{Pol}}[x]=\mathrm{tr}_\gamma \mathcal{P}\ 
e^{{i\over2}\int_0^T d\tau \sigma \omega_{\text{Pol}} }, \quad
\omega_{\text{Pol}, \mu\nu} =\frac{1}{4}( \dot x_\mu \ddot x_\nu
                               -\ddot x_\mu \dot x_\nu), \quad \dot x^2=1, 
\label{ePolSF}
\end{equation}
where $\sigma_{\mu\nu}=\frac{\I}{2} [\gamma_\mu,\gamma_\nu]$. Note
that this representation holds for propertime-parameterized
worldlines, $\dot x^2=1$. The latter property arises naturally in the
so-called ``first-order'' formalism for fermions in which the Dirac
operator acts linearly on spinor states. The Polyakov spin factor is
not only a purely geometric quantity; it also has a topological
meaning for closed worldlines. In $D=2$, it equals $(-1)^n$, with $n$
counting the number of twists of a closed loop. Generalizations of the
Polyakov spin factor to higher dimensions reveal interesting relations
to geometric quantities
\cite{Polyakov:1988,Orland:1988xq,Korchemsky:1991bk}, such as torsion
of the worldline in $D=3$, Berry phases or the notion of a Wess-Zumino
term for a bosonic worldline path integral.

However, worldline calculations with fermions are most conveniently
performed in the ``second-order'' formalism in which Dirac-algebra
valued expressions are rewritten such that the Dirac operator always
acts quadratically on spinor states
\cite{Schmidt:1993rk,Schubert:2001he}.\footnote{For a detailed
calculation in the first-order formalism, see, e.g.,
\cite{Fosco:2003rr}.}  The main advantage is that (at least for the
symmetric part of the spectrum) spinorial properties always occur in
the form of explicit spin-field couplings, such as the Pauli term
$\sim \sigma_{\mu\nu}F_{\mu\nu}$ in QED. The natural question arises
as to whether the second-order formalism can also be supplemented with
a spin-factor calculus, whether such a spin factor also has a
topological meaning and whether it opens the door to new calculational
strategies. Guided by the idea that gauge-field information can solely
be covered by a description in terms of holonomies (Wilson loops), the
existence of a spin factor can be anticipated.

In the present work, we derive such a spin-factor representation for
the second-order formalism, employing the loop-space approach to gauge
theory \cite{Polyakov:1980ca,Migdal:1984gj,Reinhardt:2000} (this
approach has recently witnessed a revival as an alternative strategy
for quantizing gravity \cite{Ashtekar:1986yd}). Concentrating on QED,
we are able to rewrite the Pauli term as a geometric quantity, i.e.,
an insertion term that depends solely on the worldline of the
fluctuating particle itself.  We develop a spin-factor calculus for
practical computations; as a concrete example, we rederive the famous
Heisenberg-Euler effective action of QED \cite{Heisenberg:1935qt}. 

In fact, we have not been able to identify a topological content
similar to that of the first-order formalism for our spin factor. But
a new geometric conclusion emerges from our formalism: it is the
continuous but non-differentiable nature of the random worldlines that
gives rise to the coupling between spin and external fields. By
contrast, smooth worldlines, i.e., smooth trajectories of a virtual
fluctuation, would not support any coupling between spinorial degrees
of freedom and an external field. Particularly the ``zigzag'' motion
of quantum mechanical worldlines mediates spin; smooth worldlines are
indeed a set of measure zero for a quantum particle.

The search for a spin-factor representation in the second-order
formalism was initiated and advanced in a series of works
\cite{Karanikas:1999hz,Avramis:2002xf}.  Therein, it was argued that
the resulting spin factor has the same form as the Polyakov spin
factor in $D=2$, cf. Eq.~\eqref{ePolSF}. For concrete computations, an
ad hoc regularization procedure was proposed to deal with possibly
arising singularities \cite{Avramis:2002xf} and was shown to work in a
variety of nontrivial examples. As our results show unambiguously, the
spin factor in the second-order formalism is not of the form of the
classic Polyakov spin factor. It is particularly the singularity
structure of our new spin factor in combination with that of the
worldlines that dominates in the second-order formalism and gives rise
to the new geometric interpretation.

Apart from intrinsic reasons for a spin-factor formalism, our work
is also motivated by the recent development of {\em worldline
numerics} \cite{Gies:2001zp,Schmidt:2002yd}, which combines the
string-inspired worldline formalism with Monte-Carlo techniques; the
result is a powerful and efficient algorithm for computing quantum
amplitudes in general background fields that has found a variety of
applications \cite{Langfeld:2002vy,Gies:2003cv,Schmidt:2003bf}. Since
Monte-Carlo methods for computing path integrals rely on the
positivity of the action, the representation of spin by
Grassmann-valued integrals is of no use for worldline numerics. Even
though fermionic worldline algorithms can be based on the conventional
Pauli-term representations \cite{Langfeld:2002vy}, the chiral limit
becomes computationally demanding. Therefore, we expect that a spin-factor
representation offers a new route to treating massless fermions with
worldline numerics. 

In this work, we develop the spin-factor formalism by considering the
fermionic determinant, which is part of any perturbatively
renormalizable gauge-field theory with charged fermions. For
simplicity, it suffices to deal with abelian gauge fields, which keeps
the presentation more transparent.  In this case, the fermionic
determinant corresponds to the one-loop effective action for photons,
i.e., the Heisenberg-Euler effective action. In Sect.~\ref{SecSF}, we
derive the spin-factor representation within the second-order
formalism for spinor QED. We elucidate the single steps in some
detail, paying particular attention to subtleties induced by the
non-analyticity of generic worldlines. In Sect.~\ref{SecASF}, we first
develop a spin-factor calculus for performing efficient computations
with the new spin factor. We apply this calculus to a rederivation of
the Heisenberg-Euler action for constant fields. Furthermore, we
combine our spin factor with a representation of the Dirac algebra in
terms of Grassmann-valued path integrals. Finally, a nonperturbative
application is given by deriving a worldline representation for the
effective action of Heisenberg-Euler type to leading nontrivial order
in $\Nf$ (quenched approximation). We summarize our conclusions in
Sect.~\ref{SecC}.

\section{Spin-factor representation in QED}
\label{SecSF}

\subsection{QED effective action on the worldline}

Let us begin with the Euclidean one-loop effective action of QED,
corresponding to the fermionic determinant \cite{Dittrich:1985yb},
\begin{equation}
\Gamma^1_\mathrm{eff}=-\ln \det\, (-\I \fsl{D} - m) =-{1\over 2}\
\ln\det(\fsl{D}\,{}^2+m^2), 
\end{equation}
where we have assumed the absence of a spectral asymmetry of the Dirac
operator in the second step;\footnote{In QED, this holds for
parity-invariant formulations which exist in any dimension. In
general, our formalism holds for the symmetric part of the Dirac
spectrum; worldline representations for the asymmetric part have been
discussed in \cite{D'Hoker:1995bj}.} the two representations of the
determinant distinguish between first-order and second-order
formalism.  Using Schwinger's propertime method
\cite{Schwinger:1951nm} together with a path-integral representation
of the propertime transition amplitude, the second-order determinant
transforms into the worldline representation,
\begin{equation}
\Gamma^1_\mathrm{eff} ={1\over2}{1\over (4\pi
)^{D/2}}\int_0^\infty {dT\over T^{1+{D/2}}}\ e^{-m^2
T}\left\langle W_\mathrm{spin}[A] \right\rangle, \label{eGam1}
\end{equation}
where the brackets denote the expectation value with respect to a path
integral over closed worldlines,
\begin{equation}
\langle \dots \rangle = \int_{x(0)=x(T)} \mathcal D x\,
\,\dots\,\, e^{-\frac{1}{4} \int_0^T d\tau\, \dot x^2(\tau)}.
\label{eWorA}
\end{equation}
We emphasize that the path integral is normalized such that $\langle
1\rangle =1$. In Eq.~\eqref{eGam1}, we have introduced the
``spinorial'' Wilson loop,
\begin{equation}
W_\mathrm{spin}[A]
= 
  \mathrm{exp}\bigg[-i \oint dx_\mu A_\mu(x) \bigg]\ 
  \mathrm{tr}_\gamma \mathcal{P}\  \mathrm{exp}\bigg[{1\over2}\int_0^T 
  d\tau\ \sigma_{\mu\nu} F_{\mu\nu} \bigg], \label{spinW} 
\end{equation}
where $\mathcal{P}$ denotes path ordering with respect to the
propertime.  The first term is the standard Wilson loop, which can be
viewed as the representation of an abstract loop operator. The last
term is the spin-field coupling with the Pauli term, which is at the
center of interest in the present work. Contrary to the standard
Wilson loop, this last term is not worldline reparametrization
invariant in the present formulation. Even though this is not
essential for our investigation, a reparametrization invariant
formulation can be constructed with the aid of an {\em einbein}
formalism \cite{Brink:1976uf}; also our results given below can
straightforwardly be generalized to such an invariant formalism.
 
\subsection{Loop derivative}

The spin-field coupling can be rewritten with the aid of the
coordinate-space representation of the loop derivative (also called
the area derivative)
\cite{Polyakov:1980ca,Migdal:1984gj,Reinhardt:2000}, 
\begin{equation}
{\delta\over\delta s_{\mu\nu}(\tau)}
=\lim_{\epsilon\to 0}\int_{-\epsilon}^{\ \epsilon} d\rho \rho\
{\delta^2 \over \delta x_\mu (\tau +{\rho\over2})\delta x_\nu (\tau
  -{\rho\over2})}, \label{eLoopD}
\end{equation} 
which is analogous to the curvature tensor in the loop representation
of gauge theory. 

Let us start with an identity that is well known in the loop space
formulation of gauge theories \cite{Reinhardt:2000}, involving
analytic functions $A_\mu(x)$ and $F_{\mu\nu}(x)$, 
\begin{equation}
 e^{-i \oint dx A(x)} \ \mathrm{tr}_\gamma \mathcal{P}\
 e^{{1\over2}\int_0^T d\tau\sigma F}
= \mathrm{tr}_\gamma \mathcal{P}\ e^{{i\over2}\int_0^T d\tau
 \sigma {\delta\over\delta s(\tau)}}\ e^{-i \oint dx A(x)}. 
\label{eFid}
\end{equation} 
It is important to stress that this relation is defined on the set of
holonomy-equivalence classes of loops in coordinate space, such that
it holds also for continuous but nondifferentiable loops;
Eq.~\eqref{eFid} can furthermore be represented with discretized
worldlines with the gauge potentials sitting on the links. 
More comments are in order: a crucial ingredient of the loop
derivative is given by the $\epsilon$ limit. A nonzero contribution
arises only if the worldline derivatives produce a specific
singularity structure $\sim \dot \delta (\rho)$, such that $\int d\rho
\, \rho\, \dot \delta (\rho) = -1$. Weaker singularities or smooth
$\rho$ dependencies vanish in the limit $\epsilon\to 0$. In
Eq.~\eqref{eFid}, the required singularity structure is provided by
the worldline derivatives acting on the gauge field and the
line-integral measure. The fact that the loop derivative can be
exponentiated rests on a property of the Wilson loop, namely,
\begin{equation}
\left[ \frac{\delta}{\delta s},  \frac{\delta}{\delta s} \right] \, 
 e^{-i \oint dx A(x)} =0, \label{eComDs}
\end{equation}
which holds only for the class of so-called Stokes-type
functionals, as introduced in \cite{Migdal:1984gj}. 

Finally, the proof of Eq.~\eqref{eFid} (as well as that of
Eq.~\eqref{eComDs}) requires a few smoothness assumptions for
worldline-dependent expressions. Whether or not they are satisfied is
a priori far from obvious, since we need these identities within the
worldline integral, but the worldlines are generically continuous but
non-differen\-tiable. This question can most suitably be analyzed with
the aid of the worldline Green's function, which reads
\cite{Schmidt:1993rk,Schubert:2001he}:
\begin{equation}
\langle x_\mu(\tau_2) x_\nu(\tau_1) \rangle \equiv -\delta_{\mu\nu}\,
G(\tau_2,\tau_1), \quad \text{with} \,\,  
G(\tau_2,\tau_1) = |\tau_2-\tau_1|-
\frac{(\tau_2-\tau_1)^2}{T}. \label{eGreen}
\end{equation}
The nonanalyticity of the worldlines becomes visible in the first term
of the Green's function, involving the modulus. By Wick contraction,
the worldline integral over general functionals of $x(\tau)$ can be
reduced to (a series of) monomials of the Green's function and its
derivatives. The following derivative is of particular importance:
\begin{equation}
\langle \ddot x_\mu(\tau_2) \dot x_\nu(\tau_1) \rangle 
=2 \dot \delta (\tau_2 -\tau_1)\, \delta_{\mu\nu}, \label{eDGreen}
\end{equation}
since the singularity structure $\sim\dot \delta$ suitable for the
loop derivative occurs. Therefore, the proof that Eq.~\eqref{eFid}
also holds under the worldline integral can be completed by the
observation that all other terms occurring during the calculation do
not involve Wick contractions of the type \eqref{eDGreen}. The same
statement applies to the proof of Eq.~\eqref{eComDs}.

Let us proceed with the spin-factor derivation by performing an
infinite series of partial integrations that shifts the loop
derivatives from the Wilson loop to the worldline kinetic term,
yielding 
\begin{equation}
\left\langle W_\mathrm{spin}[A]\right\rangle 
= \int \mathcal{D}x(\tau) \left[\mathrm{tr}_\gamma
  \mathcal{P}\ e^{{i\over2}\int_0^T d\tau \sigma {\delta\over\delta
      s(\tau)}}\  e^{-\int_0^T d\tau {\dot{x}^2(\tau)\over4}}\right]\
e^{-i \oint dx A(x)}. \label{ePartD}
\end{equation} 
No surface terms appear, since the worldlines, if stretched to
infinity, have infinite kinetic action. Now the evaluation of the
derivatives has to be performed with great care. We begin with the
leading order of the exponential series,
\begin{equation}
\bigg({i\over2}\int_0^T d\tau \sigma {\delta\over {\delta
s(\tau)}}\bigg)\bigg( e^{-\int_0^T d\tau
{\dot{x}^2(\tau)\over4}}\bigg) =\bigg({i\over2}\int_0^T d\tau\ \sigma
\omega (\tau)\bigg)\bigg( e^{-\int_0^T d\tau
{\dot{x}^2(\tau)\over4}}\bigg),
\end{equation} 
where we have defined 
\begin{equation}
\omega_{\mu\nu}(\tau)
:={1\over4}\lim_{\epsilon\to 0} \int_{-\epsilon}^{\ \epsilon}d\rho
\rho\ \ddot{x}_\mu (\tau +{\rho\over2})\ddot{x}_\nu (\tau
-{\rho\over2}). \label{eODef}
\end{equation} 
It is this $\omega$ tensor that carries the information previously
encoded in the field strength tensor. The $\omega$ tensor is
significantly different from that of Polyakov's spin factor
$\omega_{\text{Pol},{\mu\nu}}\sim (\ddot{x}_\mu
\dot{x}_\nu-\ddot{x}_\nu \dot{x}_\mu)$, arising in the first-order
formalism (cf. Eq.~\eqref{ePolSF}). For instance,
$\omega_{\mu\nu}(\tau)=0$ for any smooth loop by virtue of the
$\epsilon$ limit, whereas $\omega_{\text{Pol}, {\mu\nu}}$ is generally
nonzero in this case.

It is instructive to also study the second order in the loop
derivative explicitly:\footnote{We suppress the path ordering symbol
  here; it can easily be reinstated at the end of the calculation.}
\begin{eqnarray}
&&\left({i\over2}\int d\tau \sigma {\delta\over\delta s(\tau)}\right)^2
  e^{-\frac{1}{4} \int_0^T d\tau \dot x^2} \nonumber\\
&& =\left\{ \left({i\over2}\int d\tau \sigma \omega\right)^2
  \right. \label{eSecO}\\
&&\qquad\quad\!\! \left.
-{1\over4}\int d\tau_2 d\tau_1 \sigma_{\lambda\kappa} \sigma_{\mu\nu}
  \bigg[{\delta \omega_{\mu\nu}(\tau_1)\over \delta
        s_{\lambda\kappa}(\tau_2) } 
+ \lim_{\epsilon\to 0}\ \int\limits^{\epsilon  }_{
  -\epsilon  } d\eta  \eta 
  {\delta \omega_{\mu\nu} (\tau_1)\over\delta x_\kappa
  (\tau_2 -{\eta\over2 }) } {\ddot x_\lambda (\tau_2  +
     {\scriptstyle \frac{\eta}{2}})
  }\bigg]   
\right\}  e^{-\frac{1}{4} \int_0^T d\tau \dot x^2}. \nonumber
\end{eqnarray}
Apart from the desired first term $\sim \omega^2$, we observe the
appearance of derivatives of $\omega$. The latter correspond to a
nonvanishing right-hand side of the commutator $[\delta/\delta s,
\delta/\delta s]$ acting on the kinetic action. This is in contrast to
Eq.~\eqref{eComDs}, and reveals that the kinetic action does not
belong to the class of Stokes-type functionals. Proceeding to higher
orders in the loop derivative, the result can be represented as
\begin{equation}
 \mathrm{tr}_\gamma \mathcal{P}\ e^{{i\over2}\int_0^T d\tau
     \sigma {\delta\over\delta s(\tau)}}\  e^{-\int_0^T d\tau
     {\dot{x}^2(\tau)\over4}}
= \left[\mathrm{tr}_\gamma \mathcal{P}\ e^{{i\over2}\int_0^T d\tau
     \sigma \omega }+ D[\omega]\right]
     e^{-\int_0^T d\tau {\dot{x}^2(\tau)\over4}},  
\label{eOmD}
\end{equation} 
where $D[\omega]$ is a functional of $\omega$ that collects all terms
with at least one functional derivative of $\omega$. This functional
can formally be defined by
\begin{equation}
D[\omega]:=  e^{\int_0^T d\tau {\dot{x}^2(\tau)\over4}}\,
   \mathrm{tr}_\gamma \mathcal{P}\ e^{{i\over2}\int_0^T d\tau\sigma
      {\delta\over\delta s(\tau)}}\  
   e^{-\int_0^T d\tau  {\dot{x}^2(\tau)\over4}}
   -\mathrm{tr}_\gamma \mathcal{P}\ e^{{i\over2}\int_0^T d\tau
     \sigma \omega }\label{Ddef},
\end{equation}
with the last term simply subtracting the no-derivative
terms. An explicit representation of $D[\omega]$ can be computed order
by order in a series expansion in $\omega$; the first term, for
instance, is given by the second term in the braces in
Eq.~\eqref{eSecO}. We would like to stress that $D[\omega]$ has been
missed in the literature so far, e.g., see
\cite{Karanikas:1999hz}. However, this functional is absolutely crucial
for rendering the spin-factor representation well-defined, as will be
discussed in the next section.

\subsection{Spin factor}
\label{subsecSF}

The representation of the spin information derived in Eq.~\eqref{eOmD}
seems highly problematic. Let us recall from the definition of
$\omega$ in Eq.~\eqref{eODef} that $\omega \sim \ddot x \ddot x$. Upon
insertion into the worldline integrand, Wick contractions of the form
$\langle \ddot x \ddot x \rangle$ carrying a strong singularity
structure $\sim \ddot \delta$ will necessarily appear, cf.
Eq.~\eqref{eDGreen}. Such singularities can survive the $\epsilon$
limits and potentially render the expressions ill-defined.

In fact, we will now prove that all singularities of the type $\sim
\ddot \delta$ cancel exactly against the functional $D[\omega]$
occurring in Eq.~\eqref{eOmD}. This can straightforwardly be derived
from the zero-field limit of Eq.~\eqref{ePartD} for which the
Wilson-loop expectation value is normalized to 1,
\begin{eqnarray}
1&=&\left\langle W_\mathrm{spin}[A=0]\right\rangle 
= \int \mathcal{D}x(\tau) \, \mathrm{tr}_\gamma
  \mathcal{P}\ e^{{i\over2}\int_0^T d\tau \sigma {\delta\over\delta
      s(\tau)}}\  e^{-\int_0^T d\tau {\dot{x}^2(\tau)\over4}}
  \nonumber\\ 
&=&\int \mathcal{D}x(\tau)\ \left[\mathrm{tr}_\gamma \mathcal{P}\
e^{{i\over2}\int_0^T d\tau \sigma \omega }+D[\omega] \right] \
  e^{-\int_0^T d\tau {\dot{x}^2(\tau)\over4}},
 \label{eProvS}
\end{eqnarray}
where we have used Eqs.~\eqref{eOmD} and \eqref{Ddef} in the last
step.  Even without reference to the zero-field limit, we could have
straighforwardly proven this identity by noting that
\begin{equation}
\int \mathcal{D}x(\tau)\left({\delta\over \delta x_\mu (\tau)}
  \right)^n e^{-\int_0^T d\tau  {\dot{x}^2(\tau)\over4}}=0,\quad n\geq
  1 \label{eTotD} 
\end{equation} 
vanishes as a total derivative; we recall that the pure Gau\ss ian
velocity integral is normalized to 1.

In the language of Wick contractions, we make the important
observation from Eq.~\eqref{eProvS} that $\langle D[\omega]\rangle$
corresponds to the self-contractions of the $\omega$
exponential:\footnote{To a given order, this can algebraically be 
  confirmed by direct computation; for an explicit second-order
  calculation, see Appendix A and \cite{Haemmerling:2004}.}
\begin{eqnarray}
\langle D[\omega]\rangle&=&1- \left\langle \mathrm{tr}_\gamma \mathcal{P}\
e^{{i\over2}\int_0^T d\tau \sigma \omega } \right\rangle. \nonumber
\end{eqnarray}
Representing the worldline operators $x(\tau)$ in Fourier space by
Fock-space creation and annihilation operators of Fourier modes
(cf. Eq.~\eqref{eFourx} below), the removal of self-contractions of any
expression can be implemented by normal ordering of the Fock-space
operators; thus, we arrive at 
\begin{eqnarray}
1&=& \left\langle \mathrm{tr}_\gamma \mathcal{P}\
e^{{i\over2}\int_0^T d\tau \sigma \omega } +
 D[\omega] \right\rangle 
\equiv \left\langle \mathrm{tr}_\gamma \mathcal{P}\ :
 e^{{i\over2}\int_0^T d\tau \sigma \omega } : \right\rangle ,
 \label{eNormO}
\end{eqnarray}
where the colons denote the normal-ordering prescription. This
concludes our search for a spin-factor representation in the fermionic
second-order formalism of the worldline approach. Upon insertion into
Eq.~\eqref{eGam1}, we obtain a representation of the one-loop
contribution to the effective action for spinor QED, involving the
purely geometrical spin factor,
\begin{eqnarray}
\Gamma^1_{\mathrm{eff}}[A] &=& {1\over2}{1\over (4 \pi)^{D/2}} \int_0^\infty
{dT\over T^{(1+{D/2})}}\ e^{-m^2 T}
\int \mathcal{D}x(\tau)\ e^{-\int_0^T d\tau {\dot{x}^2(\tau)\over4}}
\ e^{-i \oint dx A(x)}\ \Phi[x], \nonumber\\
&& \text{with}\,\,\,
\Phi[x]:=\mathrm{tr}_\gamma \mathcal{P}:e^{{i\over2}\int_0^T d\tau\,
  \sigma\omega(\tau)}:\, , \label{eSFR}\\
&& \text{and}\,\,\,
\omega_{\mu\nu}(\tau)
  ={1\over4}\lim_{\epsilon\to 0} \int_{-\epsilon}^{\ \epsilon}d\rho
  \rho\ \ddot{x}_\mu (\tau +{\rho\over2})\ddot{x}_\nu (\tau
  -{\rho\over2}).\label{eODef2} 
\end{eqnarray}
An obvious advantage of this representation consists in the fact that
the dependence on the external gauge field occurs solely in the form
of a Wilson loop (holonomy). An explicit spin-field coupling no longer
appears, but spin information is extracted from the geometric
properties of the worldlines themselves. Let us emphasize once more
that a non-zero spin contribution is generated only by specific
singularity structures, arising from the continuous but
non-differentiable nature of generic worldlines.

\section{Application of the spin factor}
\label{SecASF}

\subsection{Spin-factor calculus}

Next we explore the applicability of the new spin factor in concrete
QED examples.  At a first glance, the representation of the effective
action \eqref{eSFR} seems to be disadvantageous; in particular,
concrete computations may be plagued by technical difficulties
associated with normal ordering. Moreover, even perturbative
amplitudes to finite order in $A_\mu$ seemingly receive contributions
from terms with arbitrarily high products of worldline monomials:
expanding the spin-factor and Wilson-loop exponentials, we find, for
instance, terms of the form, $\langle \omega^n \dot{x} A(x)\rangle\sim
\langle(\ddot{x}\ddot{x})^n\dot{x} A(x) \rangle$, $n$ arbitrary.

Nevertheless, it can be shown that many of these apparent high-order
contributions cancel each other and that practical calculations
actually boil down to roughly the same amount of technical work as in
the standard formalism. In view of the variety of possible worldline
monomials arising from the expansion of the Wilson loop, the spin
factor and the corresponding self-contractions (hidden behind the
normal ordering), we do not attempt to give a full account of all
possible structures and cancellation mechanisms.  Instead, we will
pick out all those terms that, upon Wick contraction, lead us
back to the full result for the effective action in standard
representation. As a result, all possible other terms ultimately have
to cancel each other.

Let us start with a new operational symbol $\{...\}_\omega^{\oint A}$
that characterizes a subclass of Wick contractions: the
$\{...\}_\omega^{\oint A}$ bracket denotes the restriction that, among
the manifold contractions arising from Wick's theorem, only those
terms have to be accounted for which are {\em complete} contractions
of one $\sigma \omega$ with {\em one and the same} $\oint dx A(x)$
factor. This already excludes many Wick contractions, in particular,
those where the two $\ddot{x}$'s out of one $\omega_{\mu\nu}$ are
either self-contracted or contracted with two different objects (be it
gauge fields or other $\omega$'s).  It turns out that all these
terms of the latter type cancel each other or vanish by the $\epsilon$
limit.  Using the Schwinger-Fock gauge,
\begin{eqnarray}
A_\alpha (x(\tau))&=& {1\over2}x_\lambda(\tau) F_{\lambda\alpha
  }(0)+{1\over3 }x_\lambda(\tau)x_\sigma(\tau)\ \partial_\sigma 
  F_{\lambda\alpha }(0)+...\nonumber\\
&=&\sum_{n=0}^\infty {x^\lambda x^{\nu_1}\cdot\cdot\cdot
  x^{\nu_n}\over n! (n+2)}\
  \partial_{\nu_1}\cdot\cdot\cdot\partial_{\nu_n}F_{\lambda\alpha}\ , 
\label{eSchF}
\end{eqnarray} 
the subclass of $\{...\}_\omega^{\oint A}$ contractions
of the Wilson-loop exponential with an $\omega$ term can
straightforwardly be computed order by order in the series
\eqref{eSchF}. The resulting series is identical to the Taylor series
of the field strength tensor, which can be summed up to yield
\begin{equation}
 \left\{{i\over2 }\int d\tau \sigma\omega\ (-i)\int d\tau\ \dot{x}_\nu
   (\tau) A_\nu(x(\tau))\right\}_{\omega}^{\oint A}   
={1\over2}\int d\tau \sigma_{\mu\nu} F^{\mu\nu}(x(\tau)).
\label{13}
\end{equation}
Since the operation of Wick contractions of bosonic fields satisfies
the elementary rules of a derivation, the same holds for the
$\{...\}_\omega^{\oint A}$ symbol. With this observation (or with
straightforward combinatorics), it follows that
\begin{equation}
 \bigg\{\mathrm{tr}_\gamma \mathcal{P}\ e^{{i\over2}\int_0^T d\tau
  \sigma  \omega(\tau)}\ e^{-i \int_0^T d\tau
  \dot{x}A(x)}\bigg\}_\omega^{\oint A}
= e^{-i \int_0^T d\tau  \dot{x}A(x)}\,\,
  \mathrm{tr}_\gamma \mathcal{P}\  e^{{1\over2}\int_0^T
    d\tau\ \sigma F}. \label{eCurlB}
\end{equation} 
This tells us immediately that it is the subclass of Wick contractions
described by the $\{...\}_\omega^{\oint A}$ symbol which already gives
us back the full result for the Pauli term. The resulting recipe is:
the spin factor can only contribute if a factor $\sim \int_0^T \sigma
\omega(\tau)$ is completely Wick contracted with a factor $\sim \oint
dx A(x)$ from the Wilson loop. Since the $\omega$-independent Wick
contractions still have to be performed, the expectation value of the
``spinorial'' Wilson loop can finally be written as
\begin{equation}
\langle W_{\text{spin}} \rangle = \left \langle 
\bigg\{\mathrm{tr}_\gamma  \mathcal{P}\ e^{{i\over2}\int_0^T d\tau 
  \sigma  \omega(\tau)}\ e^{-i \int_0^T d\tau
  \dot{x}A(x)}\bigg\}_\omega^{\oint A} \right\rangle. \label{eCurlB2}
\end{equation} 
Note that this recipe also dispenses with a consideration of normal
ordering or a detailed analysis of the self-contraction terms, since
these do not contribute to the $\{...\}_\omega^{\oint A}$ bracket by
construction. Beyond its definition via partial Wick contractions, the
$\{...\}_\omega^{\oint A}$ symbol can more abstractly be used as a
projector that removes all terms generated by self-contractions of
$\omega$ or mixed contractions as specified above. As such, the
$\{...\}_\omega^{\oint A}$ symbol is a linear operator that can
formally be interchanged with the (regularized) worldline
integral. This viewpoint will be exploited below.

The spin-factor calculus developed here has a physical
interpretation: the spin factor is only operating at those space-time
points where the fluctuating particle interacts with the external
field. The spin of the fluctuation does not generate self-interactions
of the fluctuation with its own worldline, nor does spin interact
nonlocally with the external field at two different spacetime points
simultaneously.  In the following section we demonstrate the
applicability of the spin-factor calculus by rederiving the classic
Heisenberg-Euler effective action with this new formalism.

\subsection{Heisenberg-Euler action}

As a concrete example, let us compute the one-loop effective action
for a constant background field, i.e., the Heisenberg-Euler effective
action for soft photons. We describe the background field, which is
constant in space and time but otherwise arbitrary, by the gauge
potential $A_\mu =-(1/2) F_{\mu\nu}x_\nu$. As a first simplification,
we note that path ordering is irrelevant for a constant field.
Furthermore, we observe that the path integral becomes Gau\ss ian,
since both Wilson-loop as well as spin-factor exponents depend
quadratically on $x$. The propertime derivatives become diagonal in
Fourier space where the worldlines are represented as
\begin{equation}
x_\mu(\tau)=\sum_{n=-\infty}^\infty {1\over \sqrt{T}} \ a_{n\mu}\
e^{2\pi i n\tau\over T}.  \label{eFourx}
\end{equation} 
The fact that $x_\mu\in\mathbbm{R}^D$ translates into the
reality condition $a^\ast_{-n\mu}=a_{n\mu}$. In terms of the
$a_{n\mu}$ variables, the worldline integral becomes
\begin{equation}
\langle W_{\text{spin}} \rangle =
\int \mathcal{D}a\ \mathrm{tr}_\gamma \left\{
e^{-{1\over2}\sum_n a^\ast_{\mu
    n}\Big({1\over2}\left({2\pi\over T} \right)^2 n^2
  \delta_{\mu\nu}-\left({2\pi n\over T}\right) F_{\mu\nu}+
        {1\over2}\left({2\pi n\over T} \right)^2  \sigma_{\mu\nu}\
        g_n(\epsilon)\Big) a_{n\nu}} \right\}_\omega^{\oint A}, 
\label{eHE1}
\end{equation} 
with 
\begin{equation} 
g_n(\epsilon)=\left({2\pi n\over
  T}\epsilon\right)\mathrm{cos}\left({2\pi n\over
  T}\epsilon\right)-\mathrm{sin}\left({2\pi n\over T}\epsilon\right),
  \label{eDefg}
 \end{equation} 
arising from the Fourier transform of the spin factor. Here and in the
following, the limit $\epsilon \to 0$ is implicitly understood. In
Eq.~\eqref{eHE1}, we can separate off the Fourier zero mode $n=0$,
i.e., the worldline center of mass, corresponding to the spacetime
integration of the effective action. We obtain
\begin{eqnarray}\nonumber
\langle W_{\text{spin}}\rangle = \int d^D x
 \int \mathcal{D}a\ \mathrm{tr}_\gamma 
\left\{ e^{-{1\over2}\sum_n' a^\ast_{\mu n}\ M_{\mu\nu}\  a_{\nu n} }
   \right\}_{\omega}^{\oint A} 
=\int d^D x\, \mathrm{tr}_\gamma
 \left\{\mathrm{det}'{}^{-{1\over2}}\left[{M\over
 M_0}\right]\right\}_{\omega}^{\oint A},  
\end{eqnarray}
where $M$ denotes the quadratic fluctuation operator in the exponent
of Eq.~\eqref{eHE1}. The operator $M_0$ abbreviates $M$ in
the limit $F_{\mu\nu} \rightarrow 0$ and the formal
limit\footnote{These terms would vanish anyway because of the
$\{...\}_\omega^{\oint A}$ symbol.} $g_n(\epsilon)\rightarrow 0$; the
appearance of $M_0$ implements the correct normalization of the path
integral. The prime indicates the absence of the $n=0$ zero
mode. Exponentiating the determinant results in
\begin{eqnarray}
\left\{\mathrm{det}'{}^{-{1\over2}} {M\over M_0}\right\}_\omega^{\oint
A} &=& \mathrm{exp}\left[ -{1\over2} \left\{ \sum_n{}'
\,\tr_{\text{L}} \ln \left(\mathbbm{1} - 2 \left({T\over 2 \pi n }
\right)F + \sigma g_n(\epsilon)\right)\right\}_{\omega}^{\oint
A}\right] \nonumber\\ &=&\mathrm{exp}\left[
\sum_{n=1}^\infty\,\,\mathrm{tr}_{\text{L}} \sum_{m=1}^\infty {1\over
2m} \left\{\left( 2\left({T\over 2 \pi n } \right)F - \sigma
g_n(\epsilon)\right)^{2m} \right\}_{\omega}^{\oint A}\right],
\label{eDetp}
\end{eqnarray}
where we have expanded the logarithm in the last step. Now we use the
binomial sum for the term in the $\{...\}_\omega^{\oint A}$ symbol,
\begin{equation}
\{\%\}_\omega^{\oint A}:=
\left\{ \left(  2  \left({T\over 2 \pi n } \right)F -  \sigma
g_n(\epsilon)\right)^{2m}\right\}_{\omega}^{\oint A} 
=\nonumber\sum_{k=0}^{m} 
\left(\begin{array}{c}2m \\ k\end{array}\right)  
\left(2\left({T\over 2\pi n} \right) F \right)^{2m-k} 
 \left(\sigma g_n(\epsilon) \right)^k. 
\end{equation}
Here, the $\{...\}_{\omega}^{\oint A}$ symbol has by definition
removed all those terms for which at least one $\sigma g_n(\epsilon)$
term cannot be paired with an $F$ term. This reduces the upper limit
of the sum from $2m$ to $m$. Furthermore, we have used that in the
constant field case $[F,\sigma]=0$; therefore, $F$ and $\sigma$ can
be arranged in arbitrary order.  We decompose this sum further by
separating off the $k=0$ and $k=m$ terms,
\begin{equation}\label{170}
\{\%\}_\omega^{\oint A}=
  \underbrace{\left({TF\over\ \pi n} \right)^{2m}}_{\text{(I)}} \!
  +
  \underbrace{\left(\!\begin{array}{c}2m\\ m\end{array}\!\right)
    \left({T \over\ \pi n}F\sigma g_n(\epsilon)
    \right)^m }_{\text{(II)}}\!
  +
  \underbrace{\sum_{k=1}^{m-1}  
    \left(\!\begin{array}{c}2m\\ k\end{array}\!\right) 
      \left( {T F\over \pi n}    \right)^{2m-k}\!
      \left(\sigma   g_n(\epsilon)\right)^k}_{\text{(III)}}. 
\end{equation}
The first term (I) carries no spin information; this scalar part
obviously corresponds to the contribution that we would equally
encounter in scalar QED. The second term (II) represents a perfect
pairing of spin factor and field-strength contribution; it will turn
out to contain the entire spinorial information. The remaining sum (III)
has always at least one unpaired $F$ term, even for $k=m-1$. As will
be demonstrated below, this sum vanishes completely in the $\epsilon$
limit, owing to its too-weak singularity structure.  Let us now
compute the various pieces of Eq.~(\ref{170}) separately.

Let us first consider the scalar part, substituting the first term
(I) of Eq.~\eqref{170} into Eq.~\eqref{eDetp}; we take over the result
of this standard calculation from \cite{Schmidt:1993rk,Schubert:2001he},
\begin{equation}
\text{(I)}:\qquad 
  \nonumber\mathrm{exp}\left[\sum_{n=1}^\infty\,\, \mathrm{tr}_L
  \sum_{m=1}^\infty   {1 \over 2m}  \left({T F\over \pi n }
  \right)^{2m} \right] 
=\mathrm{det}^{-{1\over2}}\left({\mathrm{sin} (FT)\over FT} \right),
\label{eScP}
\end{equation} 
where the remaining determinant refers to the Lorentz structure.  For
instance, for the constant $B$ field case, this reduces to $BT/\sinh
BT$.

Next, we consider the spinor contributions in some detail; this part
of the spin-factor-based Heisenberg-Euler calculation is genuinely
new. The spinor part induced by substitution of the second term (II) of
Eq.~(\ref{170}) into Eq.~\eqref{eDetp} can be written as
\begin{equation}
\text{(II)}:\qquad \mathrm{exp}\left[   
\sum_{m=1}^\infty {(2m-1)!\over (m!)^2}\left({T\over \pi }\right)^m
\mathrm{tr}_L\left( F\sigma\right)^m \sum_{n=1}^\infty
       {g_n(\epsilon)\over n^m}\right]. \label{eIIu}
\end{equation}
Let us discuss the Fourier sum for different values of $m$, using the
definition of $g_n(\epsilon)$ in Eq.~\eqref{eDefg},
\begin{eqnarray}\label{eFsum}
S_m:
=
  \lim_{\epsilon\to 0} \sum_{n=1}^\infty {g_n(\epsilon)\over n^m}
=
 \lim_{\epsilon\to 0} \sum_{n=1}^\infty{\left(\left({2\pi
     n\epsilon\over T}\right) \mathrm{cos}{2\pi n \over
     T}\epsilon-\mathrm{sin}{2\pi n \over T}\epsilon\right)\over n^m}
 . 
\end{eqnarray} 
For $m=1$, we have
\begin{eqnarray} 
S_1
&=&
 \lim_{\epsilon\to 0}\left[ \sum_{n=1}^\infty 
        \epsilon\ \frac{d}{d\epsilon} 
              \frac{\sin{ 2\pi n \over T}\epsilon }{n}
   -\sum_{n=1}^\infty{\mathrm{sin}{2\pi n \over T}\epsilon\over n}
 \right] \nonumber\\
&=&  \lim_{\epsilon\to 0}\left[         
   \epsilon\ \frac{d}{d\epsilon} 
     \left({{\pi-{2\pi\over T}\epsilon}\over2} \right)
     -   \left({{\pi-{2\pi\over T}\epsilon}\over2} \right) \right]
 \nonumber\\
&=&- \frac{\pi}{2}.\label{eFsum2}
\end{eqnarray}
Let us stress that this nonzero contribution survives the $\epsilon$
limit, since the Fourier sum results in a nonanalytic function
(resembling a saw-tooth profile). This agrees with our general
observation that the spin information is encoded in the nonanalytic
behavior of the worldline trajectory in spacetime. 

In fact, the $m=1$ contribution is the only nonvanishing term; all
$S_m$ for $m>1$ as well as all contributions arising from term (III)
in Eq.~\eqref{170} are zero in the limit $\epsilon \to 0$, as is shown
in Appendix B. The whole spinor contribution is that of
Eq.~\eqref{eIIu}, boiling down to $\exp \left(-(T/2)\tr_{\text{L}}
[F\sigma]\right)$. The spinorial Wilson-loop expectation value
thus becomes
\begin{equation}
 \langle W_{\text{spin}}\rangle 
= 
 \int d^D x\,
 \mathrm{det}^{-{1\over2}}\left({\mathrm{sin} (FT)\over FT} \right)
 \tr_\gamma \, e^{-\frac{T}{2}\tr_{\text{L}} [F\sigma] }
=4 \int d^D x\,
 \mathrm{det}^{-{1\over2}}\left({\mathrm{tan} (FT)\over FT} \right),
\label{eWspinR}
\end{equation}
where the Dirac trace has been taken in the last step. For instance,
for a constant $B$ field, the last line reads $4 \int d^Dx\, BT/\tanh
BT$.  Inserting this final result into Eq.~\eqref{eGam1}, we arrive at
the (unrenormalized) Heisenberg-Euler action
\cite{Heisenberg:1935qt,Schwinger:1951nm},
\begin{equation}
  \Gamma^1_\mathrm{eff}[A]
= \frac{2}{(4\pi )^{D/2}}\int_0^\infty {dT\over T^{(1+D/2)}}\
  e^{-m^2 T} \, \det{}^{-1/2} \left(\frac{\tan FT}{FT}\right).
\end{equation}
We would like to stress that the present derivation of this well-known
result is independent of other standard calculational techniques, as
far as the spinor part is concerned. The spinor contribution arises
from the subtle interplay between the purely geometric spin factor and
the Wilson loop.  Non-zero contributions arise only from terms with a
particular singularity structure. Since these singularities cannot
arise from smooth worldlines, we conclude that the random zigzag
course of the worldlines is an essential ingredient for the coupling
between spin and fields.

\subsection{Spin factor with Grassmann variables}

In the standard approaches to describing fermionic degrees of freedom,
spin information is encoded in additional Grassmann-valued path
integrals. One motivation for the spin-factor representation has been
to find a purely bosonic description devoid of both an explicit
spin-field coupling and additional Grassmann variables. 

But since the latter two criteria are independent of each other, we
can combine our spin-factor representation with Grassmann variables,
in order to make use of the elegant formulation of the Dirac algebra
and the path ordering by means of anti-commuting worldline variables.

For instance, the standard worldline formulation for the one-loop
effective action of QED in terms of a Grassmannian path integral is
given by \cite{Schmidt:1993rk,Schubert:2001he}

\begin{equation}
\Gamma^1_{\mathrm{eff}}[A]={1\over2}\int_0^\infty {dT\over T}\
  e^{-m^2 T}\int_{\mathrm{p.}} \mathcal{D}x \int_{\mathrm{a.p.}}
\mathcal{D}\psi\ e^{-\int_0^T d\tau L_{\mathrm{spin}}}, 
\end{equation} 
with 
\begin{equation}
L_{\mathrm{spin}}={1\over4}\dot{x}^2+{1\over2}\psi_\mu \dot{\psi}^\mu
+i\ \dot{x}^\mu A_\mu -i\ \dot{\psi}^\mu F_{\mu\nu} \psi^\mu. 
\label{eLspin}
\end{equation}
The path integrals satisfy either periodic (p.) or anti-periodic
(a.p.) boundary conditions, depending on their statistics.  Starting
from this representation, our line of reasoning can immediately be
applied, resulting in the following new expression for the QED action:
\begin{equation}
\Gamma^1_{\mathrm{eff}}[A]=-{1\over2}\int_0^\infty {dT\over T}\ e^{-m^2
  T}\int_{\mathrm{p.}} \mathcal{D}x \int_{\mathrm{a.p.}}
\mathcal{D}\psi\ e^{-\int d\tau{\dot{x}^2\over4}}e^{-i\oint dx A}
e^{-\int d\tau {\psi \dot{\psi}\over2}} :e^{- \int d\tau\ \dot{\psi}
  \omega \psi}:\ . 
\end{equation}  
Normal ordering takes care of the removal of $\omega$
self-contractions of the spin factor, whereas the path ordering is
automatically guaranteed by the Grassmann integral. An interesting
question of this representation concerns the fate of supersymmetry.
Whereas the standard representation has a worldline supersymmetry, the
supersymmetry is not manifest in the present formulation (the
Wilson-loop exponent and the Pauli term are supersymmetric partners in
Eq.~\eqref{eLspin}).

\subsection{Nonperturbative worldline dynamics}

The derivation of nonperturbative worldline expressions is an
application where our spin-factor representation becomes highly
advantageous. So far, we have considered perturbative diagrams
involving one charged fermion loop, but no photon
fluctuations. Promoting the fermions to a Dirac spinor with $\Nf$ 
flavor components, the functional integral over photon fluctuations
becomes Gau\ss ian in leading nontrivial order in a small-$\Nf$
expansion. In scalar QED, this gauge-field integral can be done
straightforwardly, since the worldline--gauge-field coupling occurs
simply in the form of the Wilson loop, which is a bosonic-current
interaction. In the literature, the leading-order $\Nf$ expansion has
already been used in early works on worldline techniques for scalar
QED \cite{Feynman:1950ir,Affleck:1981bm}. For instance, the
nonperturbative effective action of Heisenberg-Euler type in this
approximation reads for scalar QED,
\begin{equation}
  \GQA^{\text{Scalar QED}}[A_\mu]
= \int_x \frac{1}{4e^2} F_{\mu\nu}F_{\mu\nu} 
 - \frac{\Nf}{(4\pi)^{D/2}} \int_0^\infty \frac{dT}{T^{1+D/2}} 
    \Big\langle e^{-i  \oint dx\cdot A}\,
    e^{-\frac{e^2}{2} \int_0^T d\tau_1 d\tau_2 \dot x_{1\mu}
      \Delta_{\mu\nu}(x_1,x_2) \dot x_{2\nu} } \Big\rangle,
\label{GQAScQED} 
\end{equation}
where $\langle \dots \rangle$ again represents the worldline average
as defined in Eq.~\eqref{eWorA}.  For a detailed derivation of
Eq.~\eqref{GQAScQED}, see \cite{Gies:2005sb}.  The subscript ``QA''
refers to the leading-order $\Nf$ expansion as the ``quenched
approximation'', since diagrams with further charged loops are
neglected. In Eq.~\eqref{GQAScQED}, we have abbreviated $x_{1,2}\equiv
x(\tau_{1,2})$ and employed the photon propagator,
\begin{equation}
\Delta_{\mu\nu}(x_1,x_2)
= \frac{\Gamma(\frac{D-2}{2})}{4\pi^{D/2}} 
    \left[ \frac{1+\alpha}{2} \frac{1}{|x_1-x_2|^{D-2}}
      +({\scriptstyle \frac{D}{2}}-1)(1-\alpha) \frac{(x_1-x_2)_\mu
        (x_1-x_2)_\nu}{|x_1-x_2|^D} \right],
    \label{photprop}
\end{equation}
in $D$ dimensions with gauge parameter $\alpha$. The additional
insertion term involving the photon propagator in the worldline
average corresponds to all possible internal photon lines in the
charged loop and carries the nonperturbative contribution. It can be
shown that the quenched approximation is reliable for weak external
fields, but for arbitrary values of the coupling.\footnote{In
  nonabelian gauge theories with $\Nc$ colors,  the quenched
  approximation has also been shown to hold to leading order in a
  large-$\Nc$ expansion \cite{Strominger:1980xa}.} 

Applying the strategy of the quenched approximation to spinor QED, a
further technical complication arises from the Pauli term. Even though
the photon integral remains Gau\ss ian, the worldline current becomes
Dirac-algebra valued which has to be treated with greater care
\cite{Brambilla:1997ky}, see, e.g., \cite{Alexandrou:1998ia} for
Grassmann-valued representations. At this point, our spin-factor
approach becomes elegant, since the worldline--gauge-field coupling is
reduced to the Wilson loop. The derivation of the corresponding
nonperturbative representations in spinor QED becomes identical to
scalar QED. We can immediately write down the effective action to
leading order in $\Nf$:
\begin{eqnarray}
  \GQA^{\text{Spinor QED}}[A_\mu]
&=& \int_x \frac{1}{4e^2} F_{\mu\nu}F_{\mu\nu} 
  + \frac{\Nf}{2} \frac{1}{(4\pi)^{D/2}} \int_0^\infty
  \frac{dT}{T^{1+D/2}}  
\label{GQAQED}\\
&& \qquad\times    \Big\langle e^{-i \oint dx\cdot A}\,
    e^{-\frac{e^2}{2} \int_0^T d\tau_1 d\tau_2\, \dot x_{1\mu}
      \Delta_{\mu\nu} \dot x_{2\nu} } \,
    \mathrm{tr}_\gamma \mathcal{P}\ :
    e^{{i\over2}\int_0^T d\tau \sigma \omega } : \Big\rangle.
 \nonumber
\end{eqnarray}
This representation can now serve as the basis for nonperturbative
studies of strong-coupling QED \cite{Gockeler:1997dn} in quenched
approximation along the lines proposed in \cite{Gies:2005sb}. Further
interesting versions of this nonperturbative formula may be obtained
by trading the spin factor backwards for loop derivatives, acting now
on the Wilson loop as well as the photon insertion; this will be the
subject of future work.

\section{Conclusions}
\label{SecC}

In this work, we have used the worldline approach to quantum field
theory for a study of couplings between spinors and external gauge
fields. Guided by the idea that gauge-field information can solely be
covered by holonomies (Wilson loops), we have investigated a
reformulation of the familiar Pauli term in spinorial QED. In this
instance, we have shown that the Pauli term can be re-expressed in
terms of a spin factor which is a purely geometric quantity in the
sense that it depends only on the worldline trajectory.  Our final
representation of the fermionic fluctuation determinant, i.e., the
one-loop effective action for QED, has the following form:
\begin{equation}
\Gamma^1_\mathrm{eff}[A]
={1\over2}{1\over(4\pi)^{D\over2 }} 
  \int_{0}^{T}{ dT\over T^{1 +{D\over2}}}\ e^{-m^2T} 
  \int \mathcal{D} x(\tau)\    
  e^{- \int d\tau {\dot{x}^2(\tau)\over4 } }\,
  e^{-i\oint dx A}\,
  \mathrm{tr}_\gamma \mathcal{P}:e^{{ i\over2}\int d\tau
    \sigma \omega }:\, .
\end{equation}
The last factor represents the spin factor in the fermionic
second-order formalism with $\omega=\omega[x]$ defined in
Eq.~\eqref{eODef2}. Loosely speaking, the exponent
$\sigma_{\mu\nu}\omega_{\mu\nu}[x]$ replaces the spin-field coupling
$\sim \sigma_{\mu\nu}F_{\mu\nu}$ of the standard representation of the
fermionic effective action.

The spin factor deviates in a number of aspects from the Polyakov spin
factor, occurring in the first-order formalism. These differences,
which have been missed so far in the literature
\cite{Karanikas:1999hz,Avramis:2002xf}, are rooted in the fact that
the worldlines in the two formalisms obey different velocity
distributions: in the first-order formalism, the worldlines are
propertime-parameterized, $|\dot x|=1$, whereas their velocity is
Gau\ss ian-distributed in the second order formalism. A consequence
for the spin factors is, for instance, that smooth differentiable
worldlines give zero contribution to our spin-factor exponent;
$\omega$ has only nonzero support for worldlines of ``zigzag'' shape,
inducing a particular singularity structure. By contrast, the Polyakov
spin factor is not sensitive to the analytic properties of the
worldlines; on the contrary, it has not only a geometric but also a
topological meaning (e.g., counting the twists of a worldline in
$D=2$). We have not been ably to identify a topological meaning for
our spin factor. Even if there was one, its relevance would be
unclear, since the spin factor enters the worldline integrand with a
normal-ordering prescription. As a consequence, the spin factor in
itself does not appear to have a particular meaning; only contractions
of the spin factor with other observables such as the Wilson loop in
the integrand become meaningful.

For practical perturbative calculations, we have developed a
spin-factor calculus that reduces the amount of analytical
computational steps to roughly the same amount as in the standard
approach. The main advantage of our formulation consists in the fact
that the dependence on the external gauge field occurs solely in the
form of the Wilson loop. Particularly in computer-algebraic
realizations of high-order amplitude calculations, this may lead to
algorithmic simplifications compared to the standard approach. On the
other hand, we have to mention that the isolation of all those terms
with the required singularity structure for the $\epsilon$ limit might
lead to algorithmic complications. We have demonstrated all these
aspects in the concrete example of the classic Heisenberg-Euler
effective action. 

Our spin-factor formalism becomes truly advantageous for the analysis
of nonperturbative worldline dynamics based on the small-$\Nf$
expansion, i.e., quenched approximation. Here, the spin factor
dispenses with all complications for the photon-fluctuation
integrations induced by direct spin-field couplings. We have presented
a closed-form worldline expression for the leading-order-$\Nf$
nonperturbative effective action of Heisenberg-Euler-type that can
serve as a starting point for strong-coupling investigations. 

We believe that our work paves the way to further studies of
spin-factor representations. Our techniques are, for instance,
directly applicable to diagrams with open fermionic lines, such as
propagators, etc. We expect that our approach will become particularly
powerful in the case of nonabelian gauge fields, since the gluonic
spin-field coupling can also be traded for a spin factor. Nonabelian
gauge-field dependencies will then be described only in terms of
holonomies. In this sense, our work can be viewed as a bottom-up
approach to a loop-space formulation of gauge theories
\cite{Polyakov:1980ca,Migdal:1984gj,Reinhardt:2000}.

Since our work was also motivated by the development of worldline
numerics \cite{Gies:2001zp,Schmidt:2002yd}, we have to face the
problem of a numerical implementation of our formalism.  An immediate
numerical realization seems inhibited by the normal-ordering
prescription. This requires the study of possible alternatives.  If
the nature of our spin factor turns out to be topological, it might be
possible to classify the worldlines in terms of their topological
properties.  This would facilitate the implementation of an algorithm
that performs a Monte-Carlo sampling for each individual topological
sector separately.

To summarize, we have performed a first detailed analysis of the spin
factor in the second-order formalism of QED.  We believe that this
opens the door to many further studies of the interrelation between
spin and external fields in a geometric language.

\section*{Acknowledgment}

We are grateful to G.V.~Dunne, J.~Heinonen, K.~Klingm\"uller,
K.~Langfeld, J.~Sanchez-Guillen, M.G.~Schmidt, C.~Schubert, and
R.~Vazquez for many useful discussions. This work was supported by the
Deutsche Forschungsgemeinschaft (DFG) under contract Gi 328/1-3
(Emmy-Noether program).

\begin{appendix}

\section{Singularity cancellations: an explicit example}

Here, we demonstrate by an explicit calculation to second order that
the Wick self-contractions of the spin factor cancel against the
$D[\omega]$ term defined in Eq.~\eqref{Ddef}. This cancellation also
guarantees the absence of severe singularities. To be precise, we show
explicitly that
\begin{equation}
1= \left\langle \mathrm{tr}_\gamma \mathcal{P}\
e^{{i\over2}\int_0^T d\tau \sigma \omega } +
 D[\omega] \right\rangle \label{AppAid}
\end{equation}
holds to second order (the counting of orders can formally be defined
by the number of $\sigma_{\mu\nu}$ matrices involved). First, we
observe that the zeroth order on the RHS trivially reproduces the
LHS. The first order vanishes by virtue of the Dirac trace. The
second-order calculation requires to show that (cf. Eq.~\eqref{eSecO})
\begin{equation}
\left\langle\mathcal{P} \left(\int d\tau \sigma \omega\!\right)^2
+
\mathcal{P}\!\!\int d\tau_2 d\tau_1 \sigma_{\lambda\kappa} \sigma_{\mu\nu}
  \bigg[{\delta \omega_{\mu\nu}(\tau_1)\over \delta
        s_{\lambda\kappa}(\tau_2) } 
    + \lim_{\epsilon\to 0}\ \int\limits^{\epsilon  }_{
  -\epsilon  } d\eta  \eta 
  {\delta \omega_{\mu\nu} (\tau_1)\over\delta x_\kappa
  (\tau_2 -{\eta\over2 }) } {\ddot x_\lambda (\tau_2  +
     {\scriptstyle \frac{\eta}{2}})
  }\bigg] \right\rangle=0.\label{secord}
\end{equation}
Since the cancellation will turn out to hold already for the
$\tau_1,\tau_2$ integrands, we can suppress the path-ordering symbol
in the following.  Let us first compute the derivatives of $\omega$,
beginning with
\begin{equation}
\frac{\delta \omega_{\mu\nu }(\tau_1)}{\delta s_{\lambda\kappa}(\tau_2)}
={1\over2}\ \lim_{\epsilon_1,\epsilon_2\to 0}
  \int\limits^{ \epsilon_1}_{ - \epsilon_1}
  \int\limits^{\epsilon_2  }_{ -\epsilon_2  }
  d\eta d\rho\ \rho \eta\ \delta_{\mu\lambda}\delta_{\nu\kappa}\ 
  \ddot{\delta}\left[\tau_1 + {\scriptstyle{\rho\over2 }} 
               -(\tau_2 +{ \scriptstyle {\eta\over2 
    }})\right]
  \ \ddot{\delta}\left[\tau_1 - {\scriptstyle{\rho\over2 }}
            -(\tau_2 -{ \scriptstyle {\eta\over2}
    })\right], \label{calc1}
\end{equation}
where we have already used the antisymmetry properties of
$\omega$. Furthermore, we encounter
\begin{equation}
\frac{\delta \omega_{\mu\nu }(\tau_1)}{\delta x_\kappa(\tau_2-
  {\scriptstyle {\eta\over2}})}
={1\over2}\ 
  \lim_{\epsilon\to 0}\int\limits^{ \epsilon}_{ - \epsilon} 
  d\rho \rho\ \delta_{\mu\kappa}\ddot{x}_\nu(\tau_1-
  {\scriptstyle{\rho\over2} }) \ 
  \ddot{\delta}\left[\tau_1 + {\scriptstyle{\rho\over2 }}-(\tau_2 -
  {\scriptstyle{\eta\over2}     })\right]. \label{calc2}
\end{equation}
In order to carry out the Wick contractions, we need
the worldline propagator,
\begin{equation}
\langle x_\mu (\tau_1)x_\nu (\tau_2)\rangle = 
-\delta_{\mu\nu}\ |\tau_2 -\tau_1 | 
+\delta_{\mu\nu}\ {(\tau_2 -\tau_1)^2\over T }, 
\end{equation}
and, in particular, its propertime derivative of the form
\begin{equation}
{\langle \ddot{x}_\mu (\tau_1)\ddot{x}_\nu (\tau_2)\rangle}
 =- 2\ \ddot{\delta}(\tau_1 -\tau_2)\ \delta_{\mu\nu}.\label{dddddot} 
\end{equation}
Finally, we have to compute the contraction of the first term
of Eq.~\eqref{secord}, which involves
\begin{eqnarray}
&&\frac{1}{4}  \big\langle\ddot{x}_\mu(\tau_1+{\scriptstyle {
    \rho\over2}})\ 
\ddot{x}_\nu(\tau_1-{\scriptstyle { \rho\over2}})\ 
\ddot{x}_\lambda(\tau_2+{\scriptstyle { \eta\over2}})\
\ddot{x}_\kappa(\tau_2-{\scriptstyle { \eta\over2}})\big\rangle
    \label{calc4}\\
&&=\delta_{\mu\nu} \delta_{\lambda\kappa}\  
\ddot{\delta}\left[\tau_1 +{\scriptstyle { \rho\over2}}
      -\left(\tau_1 -{\scriptstyle { \rho\over2}}\right)\right]  \ 
\ddot{\delta}\left[\tau_2 +{\scriptstyle { \eta\over2}}
      -\left(\tau_2 -{\scriptstyle { \eta\over2}}\right)\right]
\nonumber\\
&&\quad+\delta_{\nu\lambda} \delta_{\mu\kappa}\  
\ddot{\delta}\left[\tau_1 -{\scriptstyle { \rho\over2}}
      -\left(\tau_2 -{\scriptstyle { \eta\over2}}\right)\right] 
 \ \ddot{\delta}\left[\tau_1 +{\scriptstyle { \rho\over2}}
      -\left(\tau_2 -{\scriptstyle {
          \eta\over2}}\right)\right]\nonumber\\
&&\quad  
+\delta_{\mu\lambda} \delta_{\nu\kappa}\  
\ddot{\delta}\left[\tau_1 +{\scriptstyle { \rho\over2}}
      -\left(\tau_2 +{\scriptstyle { \eta\over2}}\right)\right]  \
\ddot{\delta}\left[\tau_1 -{\scriptstyle { \rho\over2}}
      -\left(\tau_2 -{\scriptstyle { \eta\over2}}\right)\right]
.\nonumber
\end{eqnarray}
Now, inserting Eqs.~\eqref{calc1} and \eqref{calc2} into the LHS of
Eq.~\eqref{secord} and performing all Wick contractions with the aid
of Eq.~\eqref{dddddot} and \eqref{calc4}, it is straightforward to
observe that Eq.~\eqref{secord} holds as an identity. Some terms
vanish because of the contraction of $\delta_{\mu\nu}$ with
$\sigma_{\mu\nu}$, such as the first term on the RHS of
Eq.~\eqref{calc4}; all remaining terms cancel each other exactly under
the parameter integrals. This verifies the identity \eqref{AppAid} to
second order which has been proved to all orders in
Subsect. \ref{subsecSF}.

\section{Explicit calculations of spinor parts}

In the following, we show that possible further spinor parts, occurring
during the calculation of the Heisenberg-Euler action, vanish, since
they do not support a sufficient nonanalyticity. 

Let us first consider the cases of $m>1$ of the sum $S_m$, defined in
Eq.~\eqref{eFsum} and appearing in the computation of term (II) in
Eq.~\eqref{eIIu}. For this, we use an integral representation of the
function $g_n(\epsilon)$ which is defined in Eq.~\eqref{eDefg},
\begin{eqnarray}\nonumber
S_m=\lim_{\epsilon\to 0}\sum_{n=1}^\infty {g_n(\epsilon)\over n^m}
=-i \lim_{\epsilon\to 0} {2 \pi^2\over T^2} \int_{-\epsilon}^{\
  \epsilon}d\rho \rho \sum_{n=1}^\infty {e^{-\left({2i\pi \rho\over
        T}\right)n}\over n^{m-2}} 
=-{\pi\over T}\lim_{\epsilon\to 0}\int_{-\epsilon}^{\
  \epsilon}d\rho\sum_{n=1}^\infty 
{e^{-\left({2i\pi \rho\over T}\right)n}\over n^{m-1}},
\end{eqnarray}
where we have integrated by parts in the last step. In the
$\epsilon\to0$ limit, any non-zero contribution requires the $n$ sum
to exhibit a $\delta (\rho) $ singularity. As shown in the main text,
this is exactly the case for the $m=1$ term. For $m\geq3$, the $n$ sum
corresponds to a poly-logarithm of degree ${m-1}\geq 2$, which is an
analytic function for $\rho\rightarrow 0$.  Hence all $m\geq 3$ terms
vanish. The $m=2$ term is more subtle. Here we encounter
\begin{eqnarray}\nonumber
\sum_{n=1}^\infty
{e^{-\left({2i\pi \rho\over T}\right)n}\over n^{1}}
=\sum_{n=1}^\infty {\mathrm{cos}\left({2\pi \rho\over T}\right)\over
  n} +i\sum_{n=1}^\infty {\mathrm{sin}\left({2\pi \rho\over
      T}\right)\over n}. 
\end{eqnarray}
The second sum is $ \sim {\pi-\rho\over2}$ and vanishes by the
$\epsilon$ limit.  The first sum can be carried out:
\begin{eqnarray}\nonumber
\sum_{n=1}^\infty {\mathrm{cos}(n{2\pi\rho\over T})\over n}
={1\over2}\ \mathrm{ln}\left({1\over 2(1-  \mathrm{cos}{2\pi\rho\over
      T} )}\right). 
\end{eqnarray}
Therefore the $\rho$ integral becomes
\begin{eqnarray}\nonumber
-{\pi\over 2T}\int_{-\epsilon}^{\ \epsilon} d\rho\ \mathrm{ln}{1\over
 2(1-\mathrm{cos}\rho)}\approx {\pi\over T}\int_{-\epsilon}^{\
 \epsilon}\! d\rho\  \mathrm{ln}\rho\rightarrow 0\, . 
\end{eqnarray}
Even though there is a nonanalyticity, the singular structure of the
integrand is not sufficient, and the integral vanishes in the
$\epsilon\rightarrow0$ limit. This proves our first statement in the
main text that $S_m$ contributes to the effective action only in the
case of $m=1$.

Finally, we discuss the remaining sum (III) of
Eq.~(\ref{170}). Similarly to the preceding discussion, a nonzero
contribution can only arise if the result of Fourier sum over $n$ is
sufficiently singular. Concentrating on the $n$ dependence, the terms
of the Fourier sum are of the form
\begin{eqnarray}\nonumber
 {1\over n^{2m-k}}\ g_n^k(\epsilon)\sim\int_{-\epsilon}^{\ \epsilon}
 d\rho \rho   {e^{in\rho}\over n^{2m-k}}, \quad  k=1,...,m-1, \quad m>1 .
\end{eqnarray}
For all $k < m$, we end up with Fourier sums of the same type as
discussed before in this appendix; all go to zero in the $\epsilon\to
0$ limit. Hence, the whole part (III) of Eq.~\eqref{170} makes no
contribution to the effective action, as claimed in the main text.

\end{appendix}

\end{document}